\documentclass[useAMS,usenatbib,usegraphicx]{mn2e}

\usepackage{amssymb,amsmath}
\usepackage{mathrsfs}
\usepackage{times}
\usepackage{setspace}
\def\aj{AJ}%
\def\apj{ApJ}%
\def\apjl{ApJ}%
\def\aap{A\&A}%
\def\aaps{A\&AS}%
\def\mnras{MNRAS}%
\def\nat{Nature}%
\def\planss{Planet.~Space~Sci.}%
\newcommand{\beq}{\begin{equation}}
\newcommand{\eeq}{\end{equation}}

\newcommand{\msun}{\ensuremath{M_\odot}}
\newcommand{\tk}{\ensuremath{t_{\rm K}}}

\begin{document}
\voffset -1.5cm

\title[Eccentricity Oscillations in Quadruple Systems]{Greatly Enhanced Eccentricity Oscillations in Quadruple Systems Composed of Two Binaries: Implications for Stars, Planets, and Transients}

\author[Pejcha et al.]{ Ond\v{r}ej Pejcha\thanks{pejcha@astronomy.ohio-state.edu}, Joe M.\ Antognini, Benjamin J.\ Shappee\thanks{NSF Graduate Fellow}, and Todd A.\ Thompson
\vspace*{6pt}\\
Department of Astronomy and Center for Cosmology and Astroparticle Physics, The Ohio State University, 140 West 18th Avenue, Columbus, OH 43210, USA}

\maketitle

\begin{abstract}
We study the orbital evolution of hierarchical quadruple systems composed of two binaries on a long mutual orbit, where each binary acts as a Kozai-Lidov (KL) perturber on the other. We find that the coupling between the two binaries qualitatively changes the behavior of their KL cycles. The binaries can experience coherent eccentricity oscillations as well as excursions to very high eccentricity that occur over a much larger fraction of the parameter space than in triple systems. For a ratio of outer to inner semi-major axes of $10$ to $20$, about $30$ to $50\%$ of equal-mass quadruples reach eccentricity $1-e < 10^{-3}$ in one of the binaries. This is about $4$ to $12$ times more than for triples with equivalent parameters. Orbital ``flips'' and collisions without previous tidal interaction are similarly enhanced in quadruples relative to triples. We argue that the frequency of evolutionary paths influenced by KL cycles is comparable in the triple and quadruple populations even though field quadruples are a factor of $\sim 5$ less frequent than triples. Additionally, quadruples might be a non-negligible source of triples and provide fundamentally new evolutionary outcomes involving close binaries, mergers, collisions, and associated transients, which occur without any fine tuning of parameters. Finally, we study the perturbations to a planetary orbit due to a distant binary and we find that the fraction of orbital flips is a factor of $3$ to $4$ higher than for the corresponding triple system given our fiducial parameters with implications for hot Jupiters and star-planet collisions. 
\end{abstract}

\section{Introduction}

The long-term orbital evolution of hierarchical triple systems of stars and planets is subject to Kozai-Lidov (KL) cycles \citep{kozai62,lidov62} that transfer the angular momentum between the inner and outer orbits. As a result, high eccentricity and close pericenter passages of the inner binary can occur and physical effects that would not be possible otherwise can significantly change the binary dynamics. For example, tidal friction removes orbital energy and decreases the semi-major axis $a$ of the inner binary, potentially explaining the existence of short-period stellar binaries \citep[e.g.][]{mazeh79,tokovinin06,wu07,fabrycky07}, hot Jupiters, and irregular satellites of planets \citep{nesvorny03}. The high eccentricities induced by KL cycles can also reduce the gravitational wave merger timescale of compact objects (\citealp{blaes02,miller02,thompson11,antonini12,naoz12a,shappee13,hamers13}; Antognini et al., in preparation). 

The phenomenology of KL cycles is usually investigated at several levels of approximation ranging from analytic equations for the secular evolution of the orbital elements to fully numerical studies. In the simplest case of the secular evolution of a test particle subject to a three-body Hamiltonian expanded to quadrupole order (hereafter TPQ limit), the KL cycles occur only if the initial mutual inclination $i$ between the inner and outer orbits is smaller than $\cos^2 i_{\rm K} \equiv 3/5$ \citep[e.g.][]{kozai62}. The test particle achieves a maximum eccentricity of $e_{\rm max} = \sqrt{1-(5/3)\cos^2 i}$. In more realistic settings, the dynamics become more chaotic and the restrictions of the simplest case, especially the maximum eccentricity, do not apply. For example, higher-order terms of the Hamiltonian give rise to the eccentric Kozai mechanism \citep{ford00,naoz11,naoz13,lithwick11,katz11}, which operates for unequal-mass inner binaries with eccentric outer orbits, giving rise to arbitrarily high eccentricities and flips of the orientation of the inner orbit. In addition, \citet{katz12} showed with direct integration that for a small fraction of triples the angular momentum of the inner binary can go from a finite value to essentially zero in one orbit, which can produce stellar collisions with no prior tidal or gravitational wave interaction. This mechanism can perhaps produce supernovae Ia by colliding two white dwarfs. Although the full parameter space exploration of KL cycles in triple systems with direct three-body integration remains to be explored, the basic principle that systems with lower initial $|\cos i|$ reach higher eccentricities is still valid.

One of the possible stable hierarchies of quadruple systems are two binaries on a mutual orbit. In these systems, each binary acts as a distant perturber inducing KL cycles on the other binary. Since the perturber is not a point mass, the evolution of such quadruple systems can in principle differ from a combination of two uncoupled three-body KL processes. KL cycles in quadruples have not been studied previously. Recent discoveries of double close eclipsing binaries with periods very close to a 3:2 ratio \citep{cagas12,kolaczkowski13} indicate that quadruples might exhibit new and unexplained dynamics that might be related to KL cycles. Moreover, a quadruple with mutual KL cycles has also been proposed as the origin of Tycho B as a surviving companion to the Tycho SN progenitor \citep{thompson12,kerzendorf12}.

In this paper, we show that for a significant part of the parameter space the evolution of two binaries on a mutual orbit does not reduce to two independent systems consisting of a binary and distant point-like perturber. As a result of more degrees of freedom and the mutual coupling, the two binaries evolve in concert and experience excursions to very high eccentricities that would not be achieved otherwise. In Section~\ref{sec:calc}, we describe our calculations. In Section~\ref{sec:comp}, we compare the evolution of quadruples to that of triple stars. In Section~\ref{sec:coll}, we study collisions in quadruple systems. In Section~\ref{sec:disc}, we summarize our results and discuss the implications of our findings for the evolution of stars and planets.

\section{Calculations}
\label{sec:calc}

We modified the $N$-body code {\tt fewbody} \citep{fregeau04} to simulate the evolution of quadruple stars. The integrator of {\tt fewbody} is based on the GNU Scientific Library \citep{galassi11}. We tested Runge-Kutta Prince-Dormand ({\tt rk8pd}) and the Bulirsch-Stoer method of Bader and Deuflhard ({\tt bsimp}) and several conditions for adaptive timestep algorithm, all of which give qualitatively similar results for a few test cases. We do all of the calculations shown here with the default adaptive timestep algorithm with {\tt rk8pd}, which is much faster and conserves energy better than {\tt bsimp}. We also perform the calculations using Kustaanheimo-Stiefel regularization \citep{heggie74,mikkola85}, which improves the treatment of close encounters and provides well-behaved long-term energy conservation.

The dynamics of the four-body problem is inherently chaotic and eventually the numerical orbits deviate from the true ones with the same initial conditions. For calculations with absolute and relative accuracies between $10^{-11}$ and $10^{-14}$ we find that the system trajectories typically start to deviate from each other after a few times $10^{4}$ orbits of either of the inner binaries. We continue the integration for longer times since we are interested in the statistical behavior of quadruple systems. We perform all our calculations with absolute and relative integration accuracy of $10^{-12}$ and we find that this was likely an unnecessarily high accuracy since the occurrence of $1-e < 10^{-3}$ (see Section~\ref{sec:comp}) is higher only by $1$ to $3\%$ for accuracy of $10^{-11}$ while it is not converged for accuracy of $10^{-10}$. We also verified our calculations by performing simulations of triple stars, where the properties of KL cycles are known and we successfully reproduce them. Our calculations conserve total angular momentum to about $10^{-9}$ and total energy to about $10^{-8}$ of the initial value for our typical relative separations ($a_{AB}/a_A \approx 10$). However, for large relative separations ($a_{AB}/a_A \gtrsim 50$) or very different masses of the components, energy conservation is worse. We monitor the energy conservation of the calculations and remove all instances when the final relative energy conservation is worse than $10^{-5}$. However, for our fiducial choice of accuracy, the constraint on energy conservation is violated by only several runs in many thousands and thus the statistics of our results are not affected.

Since the parameter space of quadruple systems composed of binaries $A$ and $B$ on a mutual orbit $AB$ is vast, we set up the calculations by specifying the masses of all four bodies ($m_A = m_1+m_2$, $m_B=m_3+m_4$), and semi-major axes ($a_A$, $a_B$, $a_{AB}$) and eccentricities ($e_A$, $e_B$, $e_{AB}$) of the three orbits. We systematically vary the angles $i_A$ and $i_B$ between the angular momenta of orbits $A$ and $B$ with respect to their mutual orbit $AB$. The remaining parameters (arguments of pericenter, longitudes of ascending node, and mean anomalies) are chosen randomly. Typically, we perform $100$ different random initializations for each pair of $(\cos i_A, \cos i_B)$. 

Due to computational limitations, we run the calculations for $300$ Kozai times $\tk$ \citep{holman97,innanen97} of system $A$ defined as
\begin{eqnarray}
\tk &=& \frac{4}{3} \left(\frac{a_{A}^3 m_{A}}{Gm_{B}^2} \right)^{1/2} \left(\frac{a_{AB} (1-e_{AB}^2)^{1/2}}{a_{A}}   \right)^3,\quad\quad\nonumber\\
  &\simeq&  1.1\times 10^{5}\,{\rm yr} \left(\frac{a_A}{20\,{\rm AU}}\right)^3 \left(\frac{m_A}{m_{B}} \right)^{1/2}\! \left(\frac{m_{B}}{2\,\msun} \right)^{-1/2}\!\! \times \nonumber\\
& &  \times \left(\frac{a_{AB}/{a_A}}{20}\right)^3(1-e_{AB}^2)^{3/2}
\label{eq:tk}
\end{eqnarray}
Higher-order effects in KL cycles become important only on longer timescales than $\tk$. For example, the secular evolution of the eccentric Kozai mechanism occurs on a timescale of $\tk/\epsilon_{\rm oct}$ \citep{lithwick11,naoz11,naoz13,katz11}, where $\epsilon_{\rm oct}$ is the relative strength of the octupole to quadrupole term in three-body Hamiltonian expansion:
\begin{equation}
\epsilon_{\rm oct} = \left(\frac{m_1-m_2}{m_1+m_2}\right) \frac{a_A}{a_{AB}}\frac{e_{AB}}{1-e_{AB}^2}.
\label{eq:oct}
\end{equation}
The eccentric Kozai mechanism disappears for equal-mass binaries or for circular outer orbits in the secular approximation. Although Equations~(\ref{eq:tk}) and (\ref{eq:oct}) are informative for triple systems, it is unclear how they generalize to the quadruple case. Nonetheless, we use $\tk$ to determine the length of our simulation runs and $\epsilon_{\rm oct}$ serves as a characteristic of the triple dynamics that we use as a basis for understanding some quadruple phenomenology we observe. To get our results in finite time with the computational resources available to us, we limit the duration of any single calculation to $1500$\,s. This limitation becomes important only for $a_{AB}/a_A \gtrsim 20$. The quadruple dynamics is different from two uncoupled three-body KL cycles on timescales shorter than the limits we impose on the duration of the calculation. Thus, our results should be very close to the ``true'' results obtained with much longer calculation times, although technically we obtain lower bounds on the true result. We run the reference triple calculations for the same number of $\tk$ as the quadruple calculations. In a number of calculations presented in this paper, we stop the integration when a certain value of eccentricity is reached or when the orbital orientation flips.

\section{Comparison with triples}
\label{sec:comp}

\begin{figure}
\centering
\includegraphics[width=0.45\textwidth]{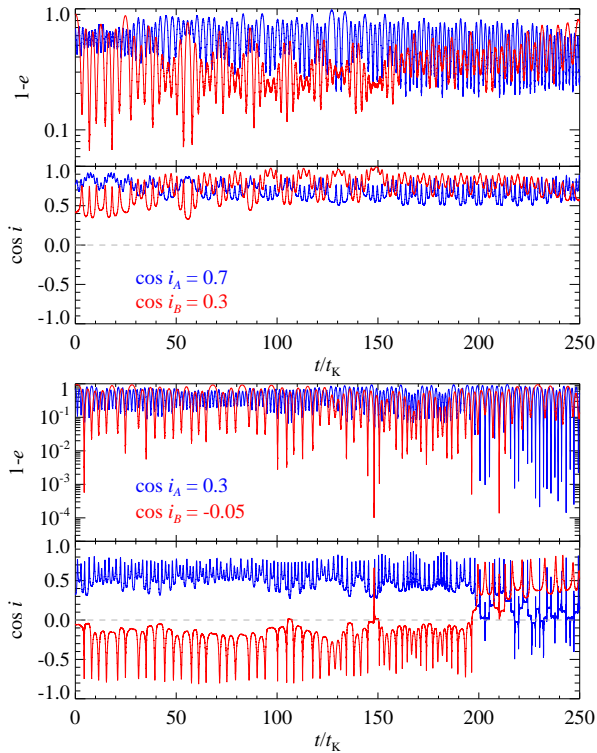}
\caption{Illustration of two possible behaviors of eccentricities $e$ and binary orientations $\cos i$ (measured with respect to the mutual orbit) in quadruple systems. The two upper panels show the two binaries evolving in concert and the two lower panels depict evolution where both binaries flip orbital orientation and go through a high-eccentricity phase. The behavior of both systems differs qualitatively from the expectation of two uncoupled triple systems. The blue lines are for system $A$ with $m_1=m_2=1\,\msun$, $a_A \approx 0.57$\,AU, and $e_A =0.3$, while the red lines are for system $B$ with $m_3=m_4=1\,\msun$, $a_B \approx 0.35$\,AU, and $e_B = 0.1$. The initial inclinations are given in the plots. The mutual orbit has $a_{AB} = 5$\,AU and $e_{AB} = 0.3$.}
\label{fig:illu}
\end{figure}

In this Section, we investigate the properties of the quadruple dynamics and we show that it is not simply a superposition of two independent KL cycles. Our fiducial calculation is for equal-mass binaries ($m_1=m_2=m_3=m_4=1\,\msun$) to eliminate the influence of eccentric Kozai mechanim for the equivalent triple systems ($\epsilon_{\rm oct} = 0$), though we consider unequal-mass binaries in Sections~\ref{sec:dep_m} and \ref{sec:dep_planet}. Summary of our quadruple calculations is given in Table~\ref{tab:sum}.

Figure~\ref{fig:illu} illustrates two examples of behavior we observe in our calculations.\footnote{We note that the evolution of the three orbits is very close to secular in the sense that the standard deviations of the three semi-major axes, and angular momentum and eccentricity of the outer orbit around the initial values are $\lesssim 0.5\%$.} The two upper panels show an initial phase of chaotic and apparently independent evolution of the two binaries followed by their long-term synchronized evolution with eccentricity modulations occurring in opposite phases. The two lower panels of Figure~\ref{fig:illu} show a qualitatively different behavior: binary $A$ initially shows only small eccentricity variations compatible with its relatively large initial inclination $\cos i_A = 0.3$, while binary $B$ reaches eccentricities $1-e_B \approx 10^{-2}$ to $10^{-3}$, which is expected from its initial inclination $\cos i_B = -0.05$. At $t\approx 148 t_{\rm K}$, binary $B$ briefly flips the direction of its orbit and reaches $1-e_B \approx 10^{-4}$. The evolution continues in a similar fashion until $t \approx 200 t_{\rm K}$ when {\em both} binaries flip their orbit orientation. As a result, binary $A$ goes through many epochs of high eccentricity ($1-e_A \approx 10^{-4}$) and the eccentricity oscillations in binary $B$ become much shallower, reaching only $1-e_B \approx 0.1$. We emphasize that in this calculation system $A$ had a moderate initial value of inclination, $\cos i_A = 0.3$, and that a triple system analogous to system $A$ with binary $B$ replaced by a point mass would not achieve such high eccentricities. In the TPQ limit ($m_1\equiv m_2$ and hence $\epsilon_{\rm oct} = 0$), the maximum eccentricity of binary $A$ would be $e_{A,{\rm max}} \approx \sqrt{1-(5/3)0.3^2} \simeq 0.92$ \citep{kozai62,innanen97,naoz13}.

\begin{figure*}
\centering
\includegraphics[width=\textwidth]{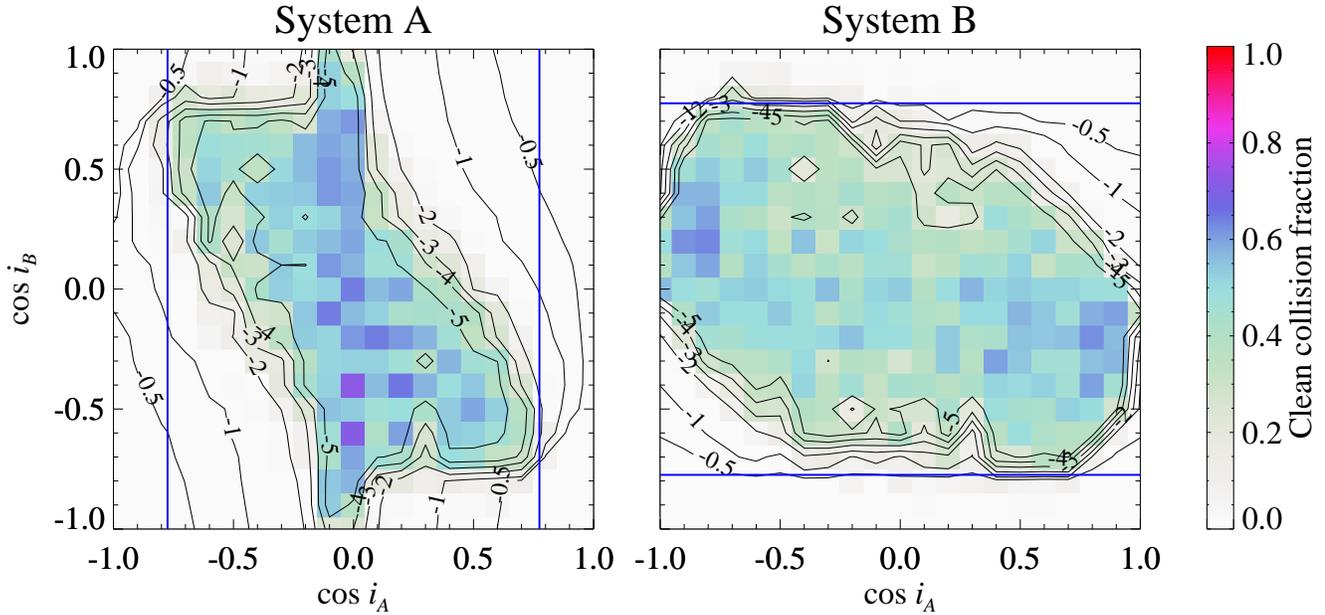}
\caption{KL cycles in quadruple systems composed of two binaries. At each pair of initial inclinations $(\cos i_A, \cos i_B)$, we performed $100$ randomly initialized calculations and recorded the highest eccentricity and the occurrence of clean collisions. The solid lines show contours of median value of the highest achieved eccentricity and are labelled by $\log_{10} (1-e)$. The color corresponds to the fraction of calculations that experienced a ``clean'' collision (pericenter distance at least four times smaller than any of the previous pericenters) as set by the colorbar. The blue horizontal and vertical lines mark the Kozai angle $\cos i_{\rm K} = \pm \sqrt{3/5}$ for each binary. The system parameters are given in Table~\ref{tab:sum} as F6.
}
\label{fig:ecc}
\end{figure*}

To characterize the quadruple properties more completely, we performed calculations spanning the full range of $(\cos i_A, \cos i_B)$. For each calculation, we recorded the highest achieved eccentricity of each binary and we show the median of these values as a function of $(\cos i_A, \cos i_B)$ using contours in Figure~\ref{fig:ecc}. The region of high eccentricities, $1-e \lesssim 10^{-3}$, extends well beyond $\cos i_A \approx 0$ and $\cos i_B \approx 0$ and thus a significant fraction of {\em both} binaries in the quadruple experience high eccentricities. In the TPQ limit, we would expect $\sqrt{(3/5)(1-0.999^2)} \simeq 3\%$ of all binaries to reach such an eccentricity. By direct integration of triple orbits we obtain about $7\%$ for $a_{AB}/a_A \approx 10$. This is still much smaller than for quadruples, where we found that $36$ and $53\%$ of quadruples reach high eccentricity in binaries $A$ and $B$, respectively (see  Section~\ref{sec:dep_a}). The region of high eccentricities is approximately bounded by the Kozai angle $\cos i_{\rm K} = \pm\sqrt{3/5}$ of each binary. The asymmetry of the contours suggests that the strongest effect is achieved when the binaries were initially on mutually retrograde orbits. The fraction of high-eccentricity systems is higher for binary $B$ than for binary $A$. The reason is that the perturbation to binary $B$ from binary $A$ and vice versa is proportional to a power of $a_A/a_{AB}$ and $a_B/a_{AB}$, respectively, and thus with the same masses and $a_{AB}$, the high-order perturbations induced by $A$ on $B$ are bigger than those of $B$ on $A$. 

Figures~\ref{fig:illu} and \ref{fig:ecc} indicate that the parameter space of high eccentricities is potentially much larger in quadruples than in triples due to the mutual coupling of the two binaries. Now we numerically explore the dependence of the coupling on semi-major axes ratios, mass ratios, and eccentricities, and compare it to triples.

\begin{table*}
\begin{minipage}{147mm}
\caption{Summary of quadruple system calculations.}
\label{tab:sum}
\begin{tabular}{lcccccccccc}
\hline
Name & $a_A$ [AU] & $e_A$ & $a_B$ [AU] & $e_B$ & $a_{AB}$ [AU] & $e_{AB}$ & $(m_1, m_2, m_3, m_4)$ [$\msun$] & $N$ & $f_A$ & $f_B$\\
\hline
F6 & 0.573 & 0.3 & 0.351 & 0.1 & 6 & 0.3 & $(1, 1, 1, 1)$ & 441 & 0.358 & 0.530\\
F12 & 0.573 & 0.3 & 0.351 & 0.1 & 12 & 0.3 & $(1, 1, 1, 1)$ & 251 & 0.146 & 0.208\\
A6 & 0.573 & 0.2 & 0.573 & 0.2 & 6 & 0.3 & $(1, 1, 1, 1)$ & 121 & 0.579 & 0.581\\
A6e0 & 0.573 & 0.2 & 0.573 & 0.2 & 6 & 0 & $(1, 1, 1, 1)$ & 121 & 0.579 & 0.581\\
A12e0 & 0.573 & 0.2 & 0.573 & 0.2 & 12 & 0 & $(1, 1, 1, 1)$ & 121 & 0.463 & 0.467\\
A24e0 & 0.573 & 0.2 & 0.573 & 0.2 & 24 & 0 & $(1, 1, 1, 1)$ & 121 & 0.244 & 0.247\\
B6 & 0.691 & 0.3 & 0.319 & 0.1 & 6 & 0.3 & $(2, 1.5, 1, 0.5)$ & 121 & 0.203 & 0.776\\
SJSS & 0.428 & 0.1 & 0.573 & 0.3 & 6 & 0.3 & $(1, 9.5\times 10^{-3}, 1, 1)$ & 441 & $-$ & $-$\\
\hline
\end{tabular}
\medskip

$N$ is the number of points in the $(\cos i_A, \cos i_B)$ plane, which might not be distributed uniformly. Each point in $(\cos i_A, \cos i_B)$ plane had $100$ calculations with randomly initialized arguments of pericenter, longitudes of ascending nodes, and mean anomalies. $f_A$ ($f_B$) is the fraction of quadruples with binary $A$ (binary $B$) reaching $1-e < 10^{-3}$. Symbols ``$-$'' indicate simulations where this quantity was not tracked. 
\end{minipage}
\end{table*}

\subsection{Dependence on semi-major axes ratio}
\label{sec:dep_a}

\begin{figure}
\centering
\includegraphics[width=0.45\textwidth]{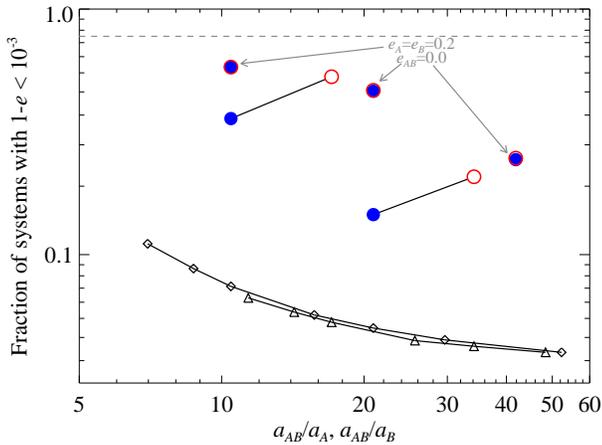}
\caption{Fraction of systems with $1-e < 10^{-3}$ as a function of the semimajor axis ratio showing that more quadruples than triples reach high eccentricity. The circles connected by a line indicate results for the two binaries $A$ (solid blue) and $B$ (open red) of quadruple systems. The triangles and diamonds connected by a line indicate results from systems where either binary $A$ or $B$ was replaced by a point mass with the same total mass. The fiducial system parameters are given in Table~\ref{tab:sum} in lines F6 and F12. 
Three overlapping circles labelled with grey arrows denote binaries with $a_A \equiv a_B$ (entries A6e0, A12e0, and A24e0 in Table~\ref{tab:sum})
The grey horizontal dashed line indicates fraction of binaries with $\cos^2 i \le \cos^2 i_{\rm K}$ assuming uniform distribution of $\cos i$.}
\label{fig:frac}
\end{figure}

We show in Figure~\ref{fig:frac} the fraction of quadruples with $1-e < 10^{-3}$ as a function of the ratio of semimajor axes $a_{AB}/a_A$ and $a_{AB}/a_B$. The orbital evolution time of our simulations is $300\tk \approx 14$\ Myr for $a_A=20$\,AU and a perturbing binary on $a_{AB}/a_A \sim 15$ orbit. We assume that the inclinations of both binaries are distributed uniformly in $\cos i$. To directly compare the quadruples with the triples, we performed a series of calculations where we replaced either of the binaries in the quadruple with a point mass with the same total mass. These triple have an equal-mass inner binary and thus the eccentric Kozai mechanism does not operate. We see from Figure~\ref{fig:frac} that for $a_{AB}/a_A \approx 10$ ($a_{AB}/a_B \approx 17$) $36\%$ ($53\%$) of quadruples reach  $1-e < 10^{-3}$ for binary $A$ ($B$). This is by a factor of $4.8$ ($10.0$) higher fraction than for otherwise equivalent triples. For wider separation of the binaries, the enhancement decreases: for $a_{AB}/a_A \approx 21$ ($a_{AB}/a_B \approx 34$) we find high-eccentricity quadruple fractions higher by a factor of $2.8$ ($4.9$) than for equivalent triples. Wider relative separations are difficult to investigate numerically, but it is likely that the fraction of quadruples experiencing high eccentricity asymptotically approaches the fraction of triples as the ratio of semimajor axes increases. Based on Figure~\ref{fig:frac}, we estimate the high-eccentricity fraction of quadruples is significantly enhanced with respect to triples for $a_{AB}/a_A \lesssim 50$. 

\subsection{Dependence on mass ratios}
\label{sec:dep_m}

So far we have discussed only quadruples composed of equal-mass stars. The strength of the interaction depends also on $q_A=m_2/m_1$, $q_B=m_4/m_3$, and $q_{AB}=m_A/m_B$. We expect that the mutual interaction vanishes for two stars orbited by a test particle ($q_A \rightarrow 0$, $q_B \rightarrow 0$), when the dynamics reduces to two independent KL cycles. To see how non-equal mass quadruples behave we ran a calculation with $q_A=3/4$, $q_B=1/2$, $q_{AB} = 7/3$ (entry B6 in Table~\ref{tab:sum}). 
As expected, the fraction of calculations where system $A$ reached high eccentricity is lower ($20\%$) than in the fiducial case because of the lower mass of the perturbing binary $m_B$. Interestingly, the fraction increases for binary $B$ ($77\%$) and essentially all systems with $\cos^2 i_B \le \cos^2 i_{\rm K} = 3/5$ reach very high eccentricity. The reason is again that $a_A/a_{AB}$ is significantly larger than $a_B/a_{AB}$, and also $m_A > m_B$. In triple systems with different inner binary masses, eccentric Kozai mechanism becomes important ($\epsilon_{\rm oct} > 0$) and induces high eccentricities over a wider range of $\cos i$ than in equal-mass systems \citep{ford00,naoz11,naoz13,lithwick11,katz11}. We verified by direct integration of triple systems that for our unequal-mass system, the fraction of quadruples reaching $1-e < 10^{-3}$ is a factor of about $3.0$ (binary $A$) to $8.2$ (binary $B$) higher than for corresponding triple systems defined by replacing one of the binaries by a point mass.

\subsection{Planets and orbital flips}
\label{sec:dep_planet}

\begin{figure}
\centering
\includegraphics[width=0.45\textwidth]{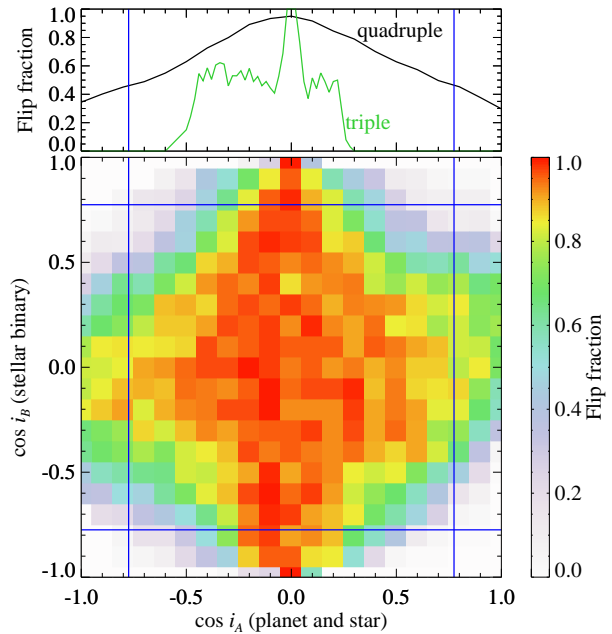}
\caption{Orbital flip fractions for star plus planet systems orbiting another binary. The lower panel shows the flip fraction as a function of the inclination of the star plus planet ($\cos i_A$) and star plus star ($\cos i_B$) binaries. The upper panel shows the flip fraction integrated over all orientations of the star+star binary (black line) in comparison to a triple system, which has the outer binary replaced by a point mass with the same total mass (green line). The system parameters are given in Table~\ref{tab:sum} as sjss.
}
\label{fig:sjss}
\end{figure}

Systems of a star and planet on a mutual orbit around a distant binary are an important example of quadruples with an extreme mass ratio. The planetary orbit can experience the eccentric Kozai mechanism leading to high eccentricity and orbital flips, which can potentially explain the abundance of observed retrograde hot Jupiters \citep{naoz11,naoz13,lithwick11,katz11}. 

In Figure~\ref{fig:sjss}, we show the flip fraction of planet orbits evaluated as a change of sign of the projection of the star plus planet angular momentum to the total angular momentum of the quadruple. For the parameters of this particular system and assuming a uniform distribution for $\cos i$ of both binaries, we find that about $66\%$ of all planetary orbits flip their orientation. How does this compare to triple systems where the distant binary is replaced by a single star? The flip fraction of triples presented in the upper panel of Figure~\ref{fig:sjss} shows clear signs of the eccentric Kozai mechanism as the broad component of the peak at $-0.5 \lesssim \cos i_A \lesssim 0.3$. Nonetheless, the total flip fraction of $\sim 21\%$ is still $\sim 3.2$ times lower than for the quadruple system. The eccentric Kozai mechanism operates on a longer timescale, $\tk/\epsilon_{\rm oct} \approx 40\tk$, for our choice of $a_A/a_{AB} \approx 0.072$ and $e_{AB} = 0.3$ and we thus ran our triple calculation for $1500\tk \approx 38\tk/\epsilon_{\rm oct}$. For our fiducial calculation length of $300\tk \approx 8\tk/\epsilon_{\rm oct}$, the flip fraction was about $2\%$ lower, suggesting that our comparison of quadruples and triples is robust.\footnote{The triple flip fraction in Figure~\ref{fig:sjss} does not look completely converged despite $100$ random initializations at each $\cos i_A$. It is possible that a significantly higher fraction of triples with $-0.5\lesssim \cos i_A \lesssim 0.3$ would eventually flip  their orbits given integration times that are orders of magnitude longer than what we are able to achieve. The flip fraction would be $\sim 40\%$ if {\em all} such orbits flipped, but this is still  $1.7$ times less than what we obtain for quadruples. It is also quite likely that the quadruple flip fraction would increase as well for very long integration times.} Although the system in Figure~\ref{fig:sjss} is not representative of typical hot Jupiter systems \citep[e.g.][]{wu07, naoz12b} and a full exploration of the parameter space is beyond the scope of this paper, our results suggest that it is much easier to place a planet on a close and possibly retrograde orbit around a star that orbits a relatively distant stellar binary. Similarly, star-planet collisions are also more likely in quadruple systems than in triples.

\subsection{Dependence on eccentricities}

Finally, the coupling between the two binaries depends also on the remaining orbital parameters that we either specified explicitly as $e_A$, $e_B$, $e_{AB}$, or marginalized over by randomly initializing the calculations (arguments of pericenter, longitudes of ascending node, mean anomalies). In order to characterize the effect of the initial eccentricities of the orbits, we performed several calculations with various choices of eccentricities and we did not find any significant changes (for example A6 and A6e0 in Table~\ref{tab:sum}), except that high initial $e_{AB}$ gives a higher fraction of quadruples that become dynamically unstable. To get a cleaner view of the effect of eccentricities and to test our code, we performed several calculations where all the explicitly specified parameters of the binaries $A$ (blue solid circle) and $B$ (red open circle) were set to identical values (entries A6e0, A12e0, and A24e0 in Table~\ref{tab:sum}) and we show the results with overlapping circles in Figure~\ref{fig:frac}. Interestingly, the fraction of quadruples composed of two identical binaries that reach $1-e < 10^{-3}$ starts decreasing considerably only for $a_{AB}/a_A \gtrsim 30$, which is later than for the fiducial case we investigated previously. Due to numerical difficulties with energy conservation at large relative separations we did not investigate this potentially interesting issue further.

\section{Collisions}
\label{sec:coll}

The KL cycles in triples can change the angular momentum of the inner binary in such a way that it goes from a finite value to essentially zero in a single orbit. As a result, stars that previously were too far from each other to interact through tides or gravitational wave emission will collide \citep{katz12}. However, for equal-mass binaries this occurs only in a few percent of systems where the perturbing body initially orbits almost perpendicularly with respect to the inner binary ($\cos i \sim 0$). In Section~\ref{sec:comp}, we showed that quadruples reach high eccentricity for much larger part of the parameter space than triples. We now investigate whether the same is true for collisions.

In a subset of our calculations, we implemented a collision detection algorithm based on \citet{katz12}. Since our calculations are dimensionless, we assume that a clean collision without any previous tidal or gravitational wave interaction occurs when a pericenter distance of the binary is at least by a factor of $4$ smaller than any of the previous ones. The eccentricity at this pericenter sets the radius of the stars relative to the semi-major axis. The upper limit on stellar radius for such ``clean'' collisions to occur is set by the time for quadruple KL cycles to develop, which scales with $\tk$, relative to the lifetime of the stars or the Hubble time. 

Color pixels in Figure~\ref{fig:ecc} show the fraction of binaries that experienced a clean collision for our fiducial calculation. Similar to the cases of eccentricity and orbital flips discussed in Section~\ref{sec:comp}, the clean collision fraction is significant over a large chunk of the parameter space and does not require any fine tuning of the inclinations. In our fiducial case displayed in Figure~\ref{fig:ecc}, about $19\%$ and $25\%$ of binaries $A$ and $B$, respectively, experience clean collisions. This is much higher than for corresponding triples, where we find $4\%$ and $2\%$, quite close to the values reported by \citet{katz12} for a similar ratio of semi-major axes. The distribution of $1-e$ at the moment of clean collisions starts at about $10^{-3}$ and peaks at about $4\times 10^{-5}$, which permits clean collisions of main-sequence and smaller stars in a Hubble time, as found also by \citet{katz12}.

\begin{figure*}
\centering
\includegraphics[width=\textwidth]{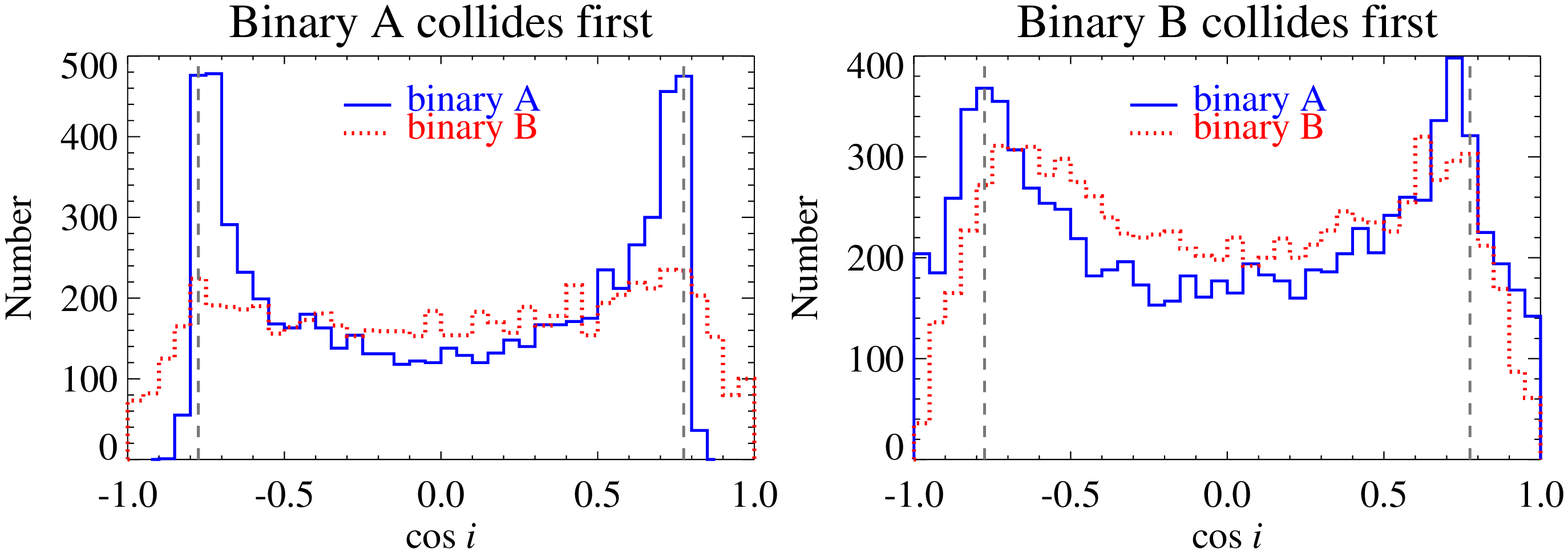}
\caption{Distribution of inclinations of the binaries with respect to their mutual orbit at the moment of first clean collision in the system for the fiducial calculations (F6 in Table~\ref{tab:sum}). Left panel shows the distribution for simulations when binary $A$ collides first, while the right panel is when $B$ collides first. Blue solid lines show binary $A$ and dotted red lines for binary $B$. Vertical gray dashed lines mark Kozai angles $\cos i_{\rm K} = \pm \sqrt{3/5}$.}
\label{fig:coll_inc}
\end{figure*}

Figure~\ref{fig:coll_inc} shows the distribution of inclinations of both binaries with respect to their mutual orbit at the moment of the first clean collision in the system. The inclination of the binary experiencing clean collision is peaked at $\cos i_{\rm K}$. Interestingly, the inclination distribution of the other binary that did not collide is also peaked at $\cos i_{\rm K}$. This effect is stronger for binary $A$, which has wider semimajor axis than binary $B$. Importantly, note that the mutual inclination of binary $B$ with respect to the tertiary (the product of the merger of the components of binary $A$) is broadly distributed.  Although a small fraction of systems ($\sim 18$\% and $23\%$ for left and right panels of Fig.~\ref{fig:coll_inc}, respectively) will not be subject to further Kozai-Lidov oscillations because $|\cos i|\leq\sqrt{3/5}$,  $77\%$ to $82\%$ will be.  Many of these systems are likely to undergo strong tidal interactions (depending on the scale of the binary), likely producing short-period binaries in triple systems, or, via collision or merger, a binary star.

\section{Discussion and Conclusions}
\label{sec:disc}

Our numerical calculations of orbital evolution of quadruple systems shows that their KL cycles are qualitatively and quantitatively different from triples. Although the binarity of the KL perturber would seem to be a higher-order effect, it significantly increases the fraction of close passages of stars in quadruple systems compared to equivalent triples. We now explore some of the implications of our findings.

We showed that high eccentricities ($1-e < 10^{-3}$), orbital flips and stellar collisions occur with a frequency $\sim 3$ to $12$ times higher in quadruples than in triples for $a_{AB}/a_A \sim 10$ to $20$ and we estimated that the frequency is still significantly higher at $a_{AB}/a_A \sim 50$. The strength of quadruple KL cycles in both binaries is maximized for equal masses. However, quadruples are also intrinsically less frequent in the field than triples. Thus, to assess the absolute occurrence of high eccentricities in the field population of quadruples and to compare it to triples, we need to know their relative frequencies, and the distributions of semi-major axes and mass ratios. The best available source is the catalog of \citet{tokovinin97,tokovinin08}. By analyzing the distance distribution of triples and quadruples composed of two binaries, \citet{tokovinin08} found that such quadruples are by a factor of $\sim 5$ less frequent than triples in the field. The median of $a_{AB}/a_A$ ($a_{AB}/a_B$) in the \citet{tokovinin08} sample is $100$ ($150$) and $\sim 40\%$ ($\sim 30\%$) of quadruples have $a_{AB}/a_A < 50$ ($a_{AB}/a_B < 50$). For triples, the median semimajor axis ratio is $\sim 80$ and $\sim 45\%$ of systems have a ratio smaller than $50$. The medians of $q_A$, $q_B$ and $q_{AB}$ of $81$ quadruple systems in the \citet{tokovinin08} sample with mass estimates are $0.64$, $0.70$, and $0.70$, respectively. To summarize, the field sample of quadruples appears to be composed of stars with similar masses and the relative distribution of semimajor axis ratios in quadruples is not significantly different from triples. Although the \citet{tokovinin08} catalog is subject to many selection effects, {\em the lower intrinsic frequency of quadruples in the field is approximately compensated for by the higher efficacy of their KL cycles}. Quadruples are thus an order unity correction to any population synthesis result involving KL cycles in triples \citep{hamers13}.

In principle, KL cycles in quadruples will act on both binaries while in a triple only the inner binary will be affected; a quadruple thus potentially produces twice as many ``interesting''  outcomes. In reality, after the first clean collision, merger, or close encounter in one of the binaries of the quadruple system, this binary will become a single star or its semimajor axis will significantly shrink due to tidal dissipation and the subsequent evolution will then proceed effectively as a triple star, with the usual properties of KL cycles in triples (see Fig.~\ref{fig:coll_inc} for distribution of inclinations at the moment of first clean collision). Assuming that the initial distribution of the ratio of semimajor axis of the inner and mutual orbits was the same in quadruples and triples, the higher efficiency of KL cycles in quadruples evolution will lead to a relative lack of quadruples with small semimajor axes ratio. This is supported by higher median of $a_{AB}/a_A$ and $a_{AB}/a_B$ in the \citet{tokovinin08} sample, however a two-way Kolmogorov-Smirnov test on the quadruple and triple semimajor axis ratio distributions does not reveal any statistically significant differences. 

The lower frequency of quadruples relative to triples in the field might also be an effect of highly efficient quadruple KL cycles: the fraction of multiple stars in star-forming regions is higher than in the field \citep[e.g.][]{leinert93,ghez97,kohler98,chen13} and the frequency of triple and quadruple stars might be comparable \citep[e.g.][]{correia06}. Thus, quadruples may be a significant source of triples and some of these triples will form binaries (see Fig.~\ref{fig:coll_inc}). Furthermore, stellar evolution and associated mass loss will lead to expansion of orbits that will be more pronounced for the inner binaries than for the mutual orbit and thus yield smaller $a_{AB}/a_A$ and $a_{AB}/a_{B}$. On the other hand, when the mass ratio of the binaries evolves farther from unity the quadruple KL cycles become weaker (Section~\ref{sec:dep_m}). Depending on how exactly the mass loss occurs, quadruple systems that did not experience KL cycles due to large separation of the binaries and unfavorable orientation of the orbit for the usual triple KL mechanism might start experiencing strong KL cycles as they evolve \citep{shappee13}. Additionally, a supernova or a neutron star kick in one member of the quadruple may also lead to interesting results \citep{shappee13}. Finally, interesting evolution in triple systems of stars and planets occurs also with KL cycles not present \citep{perets12,kratter12}.

The evolution of quadruples can produce outcomes that would not be achieved by triples and can involve main sequence stars, white dwarfs, neutron stars, and their combinations. Some outcomes include pairs of close binaries or blue stragglers on a wide orbit \citep{perets09}, blue stragglers orbiting a close binary, massive white dwarfs orbiting a (close) binary or a blue straggler, etc. Collisions and mergers of stars produce transients that can be observed over great distances. Potential examples include V838~Mon and V1309~Sco \citep[e.g.][]{soker03,tylenda06,glebbeek08,glebbeek08alt,smith11,tylenda11}. Additionally, supernovae Ia can occur as a result of a collision of two white dwarfs \citep[e.g.][]{benz89,rosswog09,raskin09,raskin10,loren10,hawley12,kushnir13}. The KL mechanism allows such collisions to happen in the field, not only in dense stellar environments, but only a few percent of triples can collide due to severe restrictions on the inclination of the perturbing body \citep{katz12}. However, we find that quadruples produce collisions generically for a large fraction of the parameter space and thus do not require fine tuning in the orientation of the orbits. Furthermore, the quadruple evolution allows for peculiar stars such as blue stragglers, close binaries, or massive white dwarfs to remain at the explosion site. Specifically, our results on KL cycles in quadruples increase by a factor of few the probability of having a metal-poor A star (blue straggler) in the Tycho supernova remnant \citep{kerzendorf12,thompson12}. The higher efficiency of quadruple KL cycles might give important constraints on the future evolution of the known quadruple systems such as those investigated by \citet{tokovinin03} and \citet{harmanec07}. Explaining the nearly 3:2 period ratios of some close double eclipsing binaries \citep{cagas12,kolaczkowski13} will require understanding the quadruple dynamics near mean-motion resonance together with tidal effects.

Finally, our results imply that stars orbiting a distant stellar binary have significantly higher probability of secular changes to their planetary orbits (Fig.~\ref{fig:sjss}) than in systems where the distant body is a point mass \citep{wu03,wu07,fabrycky07,naoz11,naoz13,katz11,lithwick11}. Without a more detailed investigation of the parameter space, it is not clear whether the enhanced KL cycles in such systems would lead to a higher fraction of hot Jupiters (possibly on retrograde orbits) or more star-planet collisions. \citet{roell12} list nine exoplanet host stars that are members of triples that are in the hierarchy considered here. For most of these systems, the separation of the host star from the distant binary relative to its separation is within the range considered in this paper, $a_{AB}/a_B \lesssim 50$. Planets in these systems are not hot Jupiters, but the secular perturbation timescale would be large due to the size of the stellar orbits, which are often visually resolved. 

\section*{acknowledgements}
We thank Christopher Kochanek, Smadar Naoz, and Kaitlin Kratter for discussions and a detailed reading the manuscript. We thank Boaz Katz and Subo Dong for discussions. We thank our referee, Will Farr, for constructive comments that helped to improve the paper. We thank J.\ Fregeau for making the code {\tt fewbody} publicly available. O.P.\ thanks Andy Gould, and the members of Princeton University Department of Astrophysical Sciences and Institute for Advanced Study for discussions of KL cycles and the enigmatic object CzeV343. B.J.S.\ was supported by a Graduate Research Fellowship from the National Science Foundation.

\end{document}